\documentclass[lettersize,journal]{IEEEtran}
\usepackage{amsmath,amsfonts}
\usepackage{algorithmic}
\usepackage{algorithm}
\usepackage{array}
\usepackage[caption=false,font=normalsize,labelfont=sf,textfont=sf]{subfig}
\usepackage{textcomp}
\usepackage{stfloats}
\usepackage{url}
\usepackage{verbatim}
\usepackage{graphicx}
\usepackage{cite}
\usepackage[colorlinks,
linkcolor=blue,
anchorcolor=blue,
citecolor=blue,backref]{hyperref}
\usepackage[all]{hypcap}
\usepackage{multirow}
\usepackage{ulem}
\usepackage{float}
\begin{document}

\title{3-D Magnetotelluric Deep Learning Inversion Guided by Pseudo-Physical Information}

\author{Peifan Jiang, Xuben Wang, Shuang Wang, Fei Deng, Kunpeng Wang, Bin Wang, and Yuhan Yang\vspace{-5mm}

\thanks{This work is supported by the National Natural Science Foundation of China under Grant 41930112. {(\textit{Corresponding author: Xuben Wang and Fei Deng.})}}

\thanks{Peifan Jiang, Xuben Wang, Shuang Wang, Kunpeng Wang and Yuhan Yang are with the Key Laboratory of Earth Exploration and Information Techniques of Ministry of Education, College of Geophysics, Chengdu University of Technology, Chengdu 610059, China (e-mail: jpeifan@qq.com; wxb@cdut.edu.cn).

Fei Deng and Bin Wang are with the College of Computer Science and Cyber Security, Chengdu University of Technology, Chengdu 610059, China (e-mail: dengfei@cdut.edu.cn).

}}

\markboth{}%
{Shell \MakeLowercase{\textit{et al.}}: A Sample Article Using IEEEtran.cls for IEEE Journals}


\maketitle

\begin{abstract}
Magnetotelluric deep learning (DL) inversion methods based on joint data-driven and physics-driven have become a hot topic in recent years. When mapping observation data (or forward modeling data) to the resistivity model using neural networks (NNs), incorporating the error (loss) term of the inversion resistivity's forward modeling response—which introduces physical information about electromagnetic field propagation—can significantly enhance the inversion accuracy. To efficiently achieve data-physical dual-driven MT deep learning inversion for large-scale 3-D MT data, we propose using DL forward modeling networks to compute this portion of the loss. This approach introduces pseudo-physical information through the forward modeling of NN simulation, further guiding the inversion network fitting. Specifically, we first pre-train the forward modeling networks as fixed forward modeling operators, then transfer and integrate them into the inversion network training, and finally optimize the inversion network by minimizing the multinomial loss. Theoretical experimental results indicate that despite some simulation errors in DL forward modeling, the introduced pseudo-physical information still enhances inversion accuracy and significantly mitigates the overfitting problem during training. Additionally, we propose a new input mode that involves masking and adding noise to the data, simulating the field data environment of 3-D MT inversion, thereby making the method more flexible and effective for practical applications.
\end{abstract}


\section{Introduction}
\IEEEPARstart{T}{he} Magnetotelluric (MT) method is a crucial technology in geophysical exploration. The technique is widely applied in engineering fields like mineral exploration\cite{strangway1973application}~\cite{farquharson2009three}, geothermal exploration\cite{xue2023three}~\cite{newman2008three}, and oil and gas\cite{vozoff1991magnetotelluric}~\cite{vozoff1972magnetotelluric} extraction by receiving and analyzing natural electromagnetic field signals to inversion and image subsurface structures. Moreover, the MT method is an essential observational tool for deep Earth science research and the analysis of plate tectonic dynamics\cite{liu2016high}~\cite{akinremi2022crustal}~\cite{unsworth2010magnetotelluric}.

Traditional MT inversion algorithms are mainly classified into two categories: locally optimized deterministic inversion and stochastic inversion with global optimization capabilities. Locally optimized inversion methods include the nonlinear conjugate gradient (NLCG)\cite{newman2000three}, Occam inversion\cite{constable1987occam}~\cite{siripunvaraporn2000efficient}, Gauss-Newton\cite{nadasi2022large} and limited-memory quasi-Newton\cite{avdeev20093d} methods. These methods rely on the initial model and often converge to local optimization solutions, which can result in overfitting. On the other hand, stochastic inversion includes simulated annealing algorithm\cite{sen1993nonlinear}, genetic algorithm\cite{boschetti2001interactive}, particle swarm optimization\cite{shaw2007particle}, and Bayesian approach\cite{xiang2018efficient}. Although these methods can effectively achieve global optimization, but their application in large-scale 3D MT inversion is still challenging due to the high demand of computational resources and relatively low efficiency. MT inversion requires iterative invocation of forward modeling. By continuously updating the subsurface resistivity model, the forward modeling response is gradually approximated to the observed data, thereby enabling the imaging of subsurface structures. MT forward modeling is based on Maxwell's electromagnetic equations and is typically implemented using numerical computation methods such as the finite element method (FEM)\cite{FEM2}, finite difference method (FDM)\cite{FDM1}, finite volume method (FVM)\cite{FVM1}, and integral equation method\cite{equationM}. Different forward modeling algorithms have distinct advantages and disadvantages, which are applicable to different geological conditions and computational needs.

With the booming development of artificial intelligence (AI) in various fields, deep learning (DL) methods using neural networks (NNs) to simulate the forward and inverse numerical computations of conventional electromagnetic fields have garnered significant attention\cite{guo2021physics}~\cite{li2018deepnis}~\cite{puzyrev2019deep}. For DL forward modeling, Shan \textit{et al.} approached 2-D MT forward modeling as a multi-task learning problem, synchronizing the prediction of apparent resistivity and phase using NNs\cite{Shan2021TE}. Deng \textit{et al.} similarly implemented 2-D MT forward modeling by multiple NNs and expanded the dataset by traditional iterative inversion to improve the forward modeling accuracy and generalization\cite{Deng2023MT2D}~\cite{Deng2023MT2D2}. Additionally, Wang \textit{et al.} utilized NNs for 3-D MT DL forward modeling, achieving accurate forward modeling response predictions of multimorphic anomalies in 3-D space\cite{wang2024three}. For DL inversion, most researchers employed NNs to directly establish the nonlinear mapping relationship from observed data to geoelectric model, exploring new inversion paradigm. Among them, Liu \textit{et al.} proposed using convolutional neural network (CNN) for end-to-end electrical resistivity inversion\cite{liu2020deep}. To address the inherent vertical variation in apparent resistivity data, they incorporated additional layer feature mapping to include depth information, achieving improved inversion results. Ling \textit{et al.} proposed using ResNet1D for audio MT inversion, enhancing network performance by adding shortcut connections, and achieving robust end-to-end inversion\cite{ling2023one}. Liao \textit{et al.} proposed a fully convolutional network (FCN) -based MT inversion method to directly establish the mapping of apparent resistivity and phase to resistivity anomaly models, and discussed the ability of the network can be trained in a single anomaly model dataset and generalized to multiple anomaly models due to the weight-sharing mechanism present in FCN\cite{liao20222d}. Addressing the multi-scale characteristics of MT forward modeling response data, Pan et al. proposed MT2DInv-Unet\cite{pan2024mt2dinv}, a deformable convolution-based 2-D MT inversion network that outperforms both FCN and U-Net in MT inversion. Xu et al. introduced the self-attention mechanism to enhance the feature acquisition capabilities of traditional CNNs and designed more representative training samples to improve inversion effectiveness\cite{xu2024magnetotelluric}.

The above DL methods focus solely on the mapping error between observed data and subsurface resistivity during training, neglecting forward modeling calculations and the underlying physical relationships. As a result, significant effective information is lost, limiting the accuracy and generalization of the inversion. In recent years, the integration of physical information and guidance with physics-driven inversion methods has garnered significant attention. Alyousuf \textit{et al.} employed an adaptive physical embedding approach to combine DL inversion and deterministic inversion, introducing the loss of the geoelectric model term via NN inversion during the iterative process of the deterministic inversion to achieve 2-D MT inversion\cite{alyousuf2022inversion}. Other researchers used forward modeling algorithm as an additional constraint for NN training, incorporating the physical information of electromagnetic field propagation into DL inversion training through forward modeling. Liu \textit{et al.} employed traditional forward modeling method to compute the forward response of the predicted resistivity of the network based on the construction of a nonlinear mapping from apparent resistivity to resistivity\cite{liu2023physics}. They incorporated physical information by using the difference between the forward modeling response and the input apparent resistivity as part of the loss function. Similarly, Liu \textit{et al.} and Ling \textit{et al.} embedded traditional forward modeling algorithms into 1-D and 3-D MT DL inversion training, reducing the network's overdependence on the training set and enhancing the stability and accuracy of inversion\cite{liu2022physics}~\cite{ling2024three}. These studies demonstrate that incorporating physical information (forward modeling computation) into network training effectively improves inversion accuracy, but the challenges are also obvious. In general, achieving DL inversion with stronger generalization ability and accuracy still requires training on large-scale datasets, even when guided by physical information. Therefore, when applying this method to train large-scale 3-D MT data, involving tens or even hundreds of thousands of grids, the time cost of traditional forward modeling computations becomes prohibitive during network training, severely limiting the method's applicability.

To address the above issues, we combine 3-D DL forward modeling with 3-D DL inversion, and introduce pseudo-physical information into the inversion network training through NN-based forward simulation, so as to propose a 3-D MT DL inversion method guided by pseudo-physical information. The main research is as follows:
\subsubsection*{\bf 1)}
Incorporate pseudo-physical information into 3-D MT DL inversion training. We utilize pre-trained DL forward modeling sub-networks as fixed forward modeling operators to rapidly compute the 3-D MT forward modeling response. During DL inversion training, the forward modeling sub-networks compute the forward response error of the inversion resistivity, which further guides network updates. Experimental results indicate that, compared with the direct training inversion (DirectInv), which does not consider forward modeling response errors, the P-PhysInv method improves the inversion accuracy and significantly reduces network overfitting.

\subsubsection*{\bf 2)}
Construct a 3-D MT deep learning theoretical dataset with layered structures. By interpolating surfaces of random points in 3-D space, we generate formation surfaces with multimorphic layered structures and fill the random resistivity between adjacent surfaces. Meanwhile, we add several regular anomalies within the layers to construct a complex geoelectric model that reflects actual subsurface structures more accurately. Finally, we use ModEM to perform forward modeling computation via FDM, generating sample datasets for training the forward modeling and inversion networks.

\subsubsection*{\bf 3)}
Propose a data input mode with more practical application potential. During the training process, we performed missing (i.e., the mask process) and adding noise to the input data (forward modeling data) to simulate a sparse and noisy real data environment. This mode not only enables the network to better learn the features of each apparent parameters, but also enhances the method’s flexibility for application to field data. Experiments show that for theoretical data, the network can effectively generalize to data with higher noise and mask levels after training at relatively low noise and mask levels. For field data, we explore the effectiveness of this mode for end-to-end 3-D MT DL inversion in sparse data conditions, and discuss the possible further enhancement effect of using the inversion results as an initial model for traditional inversion.

\section{Methodological}
\subsection{Traditional deterministic inversion}

\begin{figure*}
	\centering
	\includegraphics[width=1\linewidth]{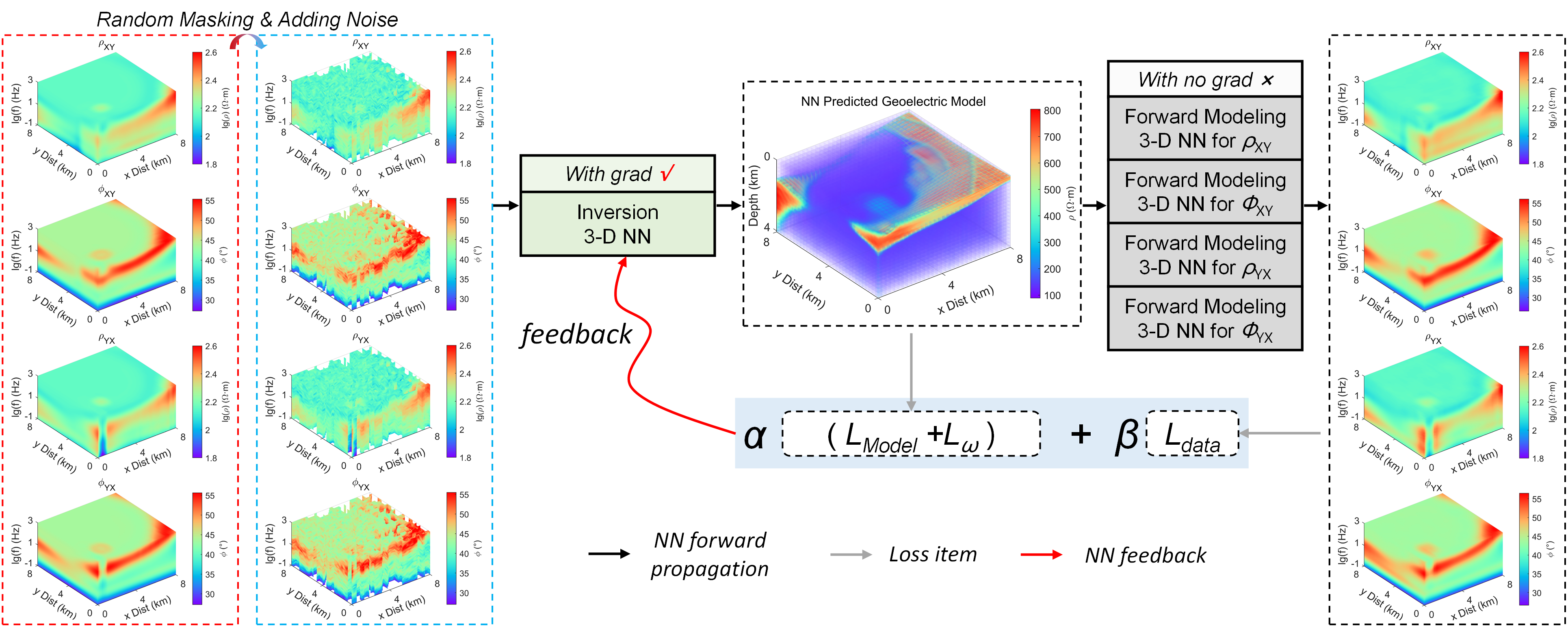}
	\caption{Network Training Process for P-PhysInv}
	\label{fig:fig1}
\end{figure*}

The target function ${\Phi _t}$ of the traditional deterministic 3-D MT inversion method mainly consists of Data Misfit Term ${\Phi _{dm}}$ and Model Regularization Term ${\Phi _{mr}}$, defined as:
\begin{equation}
	{\Phi _t} = {\Phi _{dm}} + \lambda {\Phi _{mr}}
	\label{equ1}
\end{equation}
where $\lambda$ is a weighting factor to balance the data fitting term and the model constraint term. ${\Phi _{dm}}$, ${\Phi _{mr}}$ are defined as:
\begin{equation}
	\begin{array}{c}
		{\Phi _{dm}} = {\left( {Jm - {d_{obs}}} \right)^T}W_d^T{W_d}\left( {Jm - {d_{obs}}} \right)\\
		= {\left\| {{W_d}\left( {Jm - {d_{obs}}} \right)} \right\|^2}
	\end{array}
		\label{equ2}
\end{equation}
\begin{equation}
	\begin{array}{c}
		{\Phi _{mr}} = {\left( {m - {m_{ref}}} \right)^T}W_m^T{W_m}\left( {m - {m_{ref}}} \right)\\
		= {\left\| {{W_m}\left( {m - {m_{ref}}} \right)} \right\|^2}
	\end{array}
		\label{equ3}
\end{equation}
where ${d_{obs}}$ denotes observed data; $Jm$ denotes predicted observed data, derived by solving the Jacobian matrix $J$ for the target geoelectric model $m$; ${m_{ref}}$ denotes the reference model in the model regularization term; and $W_d$ and $W_m$ denote the weight matrices of the different terms, respectively.

According to Equation (\ref{equ1}), (\ref{equ2}), (\ref{equ3}), traditional deterministic inversion methods primarily focus on minimizing the data fitting term, typically involving the solution of the Jacobian matrix to derive the forward modeling response of the target geoelectric model. In an optimized inversion framework, forward modeling is continuously invoked during the inversion process, and the fitting difference between the forward modeling data and the observed data guides the updating of the geoelectric model. When the fitting difference reaches a certain threshold or the inversion algorithm completes a preset number of iterations, the inversion is terminated, yielding the target geoelectric model as the final result. In this process, the direct mapping from observed data to subsurface resistivity is not established.

\subsection{DL inversion and forward modeling}
The difference with traditional methods is that common DL inversion methods directly map the data to the model, \textit{i.e.}, establish a mapping relationship between $d_t$ (observed data / forward modeling data) and $m_l$ (predicted geoelectric model). The nonlinear relationship between the two is represented using a neural network as:
\begin{equation}
	{m_l} = {f^T}[{d_t};\theta ]
\end{equation}
where $\theta$ denotes the hyperparameters of the network (including weights, biases, \textit{etc.}); ${f^T}$ is a pseudoinverse operator parameterized by $\theta$\cite{guo2004pseudoinverse}. Thereafter, the exact mapping from $d_t$ to $m_l$ will be achieved by minimizing the objective function (loss function) ${L_{inv}}$, to narrow the gap between the predicted geoelectric model $m_l$ and the target geoelectric model $m_t$. The process can be expressed as:
\begin{equation}
	{L_{inv}} = {\min _\theta }\left\| {{m_l} - {m_t}} \right\|_{{L_2}}^2 = {\min _\theta }\left\| {{f^T}[{d_t};\theta ] - {m_t}} \right\|_{{L_2}}^2
	\label{equ5} 
\end{equation}

Similarly, the nonlinear relationship of MT DL forward modeling using NNs can be expressed as:
\begin{equation}
	{d_l} = {f^T}[{m_t};\theta ]
	\label{equ6}
\end{equation}
where if $d$ is denoted as different observed data or forward modeling data (apparent resistivity and phase for different polarization directions), we can then make the network predict the correct forward modeling results by minimizing multiple loss functions ${L_{fm}}$ as follows:
\begin{equation}		
	\sum {{L_{fm}}}  = \sum {{{\min }_\theta }\left\| {{d_l} - {d_t}} \right\|_{{L_2}}^2}  = \sum {{{\min }_\theta }\left\| {{f^T}[{m_t};\theta ] - {d_t}} \right\|_{{L_2}}^2}	
	\label{equ7} 
\end{equation}

\subsection{DL inversion combined with DL forward modeling}

Therefore, if we consider the pretrained forward modeling network model as a fixed forward operator, it can be used to calculate the forward response of the subsurface resistivity predicted by the inversion network, and thus to calculate the model-to-data error. According to Equation (\ref{equ5}) and (\ref{equ7}), the objective function of DL inversion combined with DL forward modeling can be expressed as:
\begin{equation}
	{L_{obt}}(m,\theta ) = \alpha {L_{inv}} + \beta \sum {{L_{fm}}}
		\label{equ8}  
\end{equation}
where $\alpha $ and  $\beta $ are the weights of different terms. Different from previous studies, the construction of this objective function is based on NNs in the whole process, and we construct the data-to-model mapping A and the model-to-data mapping B through NN, respectively. Among them, ${L_{inv}}$ corresponding to mapping A serves as the main fitting target aiming at realizing the inversion, and ${L_{fm}}$ corresponding to mapping B can be regarded as constraints and guide terms to provide forward modeling information with physical meanings for the network. Compared with using traditional forward modeling algorithms, DL forward modeling greatly improves the efficiency and allows us to incorporate physical information into large-scale 3-D MT inversion training.

To achieve this goal, we use a multi-stage training approach. We first pretrain multiple forward modeling sub-networks and consider them as fixed forward modules, and then combine these forward modeling modules to train inversion sub-network. Specifically, in the first stage, we follow the forward modeling network training strategy in the literature [1] and use multiple forward modeling sub-networks to train the forward modeling process for different forward response data (including  ${\rho _{XY}}$, ${\phi _{XY}}$ , ${\rho _{YX}}$ and ${\phi _{YX}}$), respectively. In the second stage, we concatenate the inversion sub-network with these four pre-trained forward modeling sub-networks, the outputs of the inversion sub-network will be used as the inputs of the four forward modeling sub-networks, constituting a large network with five sub-networks. The parameter weights of the four forward modeling sub-networks will be frozen during the training process, and they will not have gradients during back-propagation, and we only update the parameters of the inversion sub-network.

In addition, in the second stage of the training process, we propose a different data input mode from the existing studies, for each input apparent parameter data we will randomly add 0-6\% Gaussian noise. At the same time, we will also randomly eliminate 0-60\% of the measurement point data (both Mask operation), it should be noted that there will be no aggregation mask as well as we will ensure the integrity of the central measurement line. This input mode can be regarded as targeted data enhancement, which not only facilitates the network to improve the analytical capability of target characterization, but also conforms to the situation of noise interference and sparse distribution of observed data in the actual 3-D MT exploration, and improves the flexibility of the practical application of the method. The training process is shown in Figure \ref{fig:fig1}, where we will guide the network optimization by the total loss function composed of multiple loss functions to fit the inversion sub-network.

\begin{figure}
	\centering
	\includegraphics[width=1\linewidth]{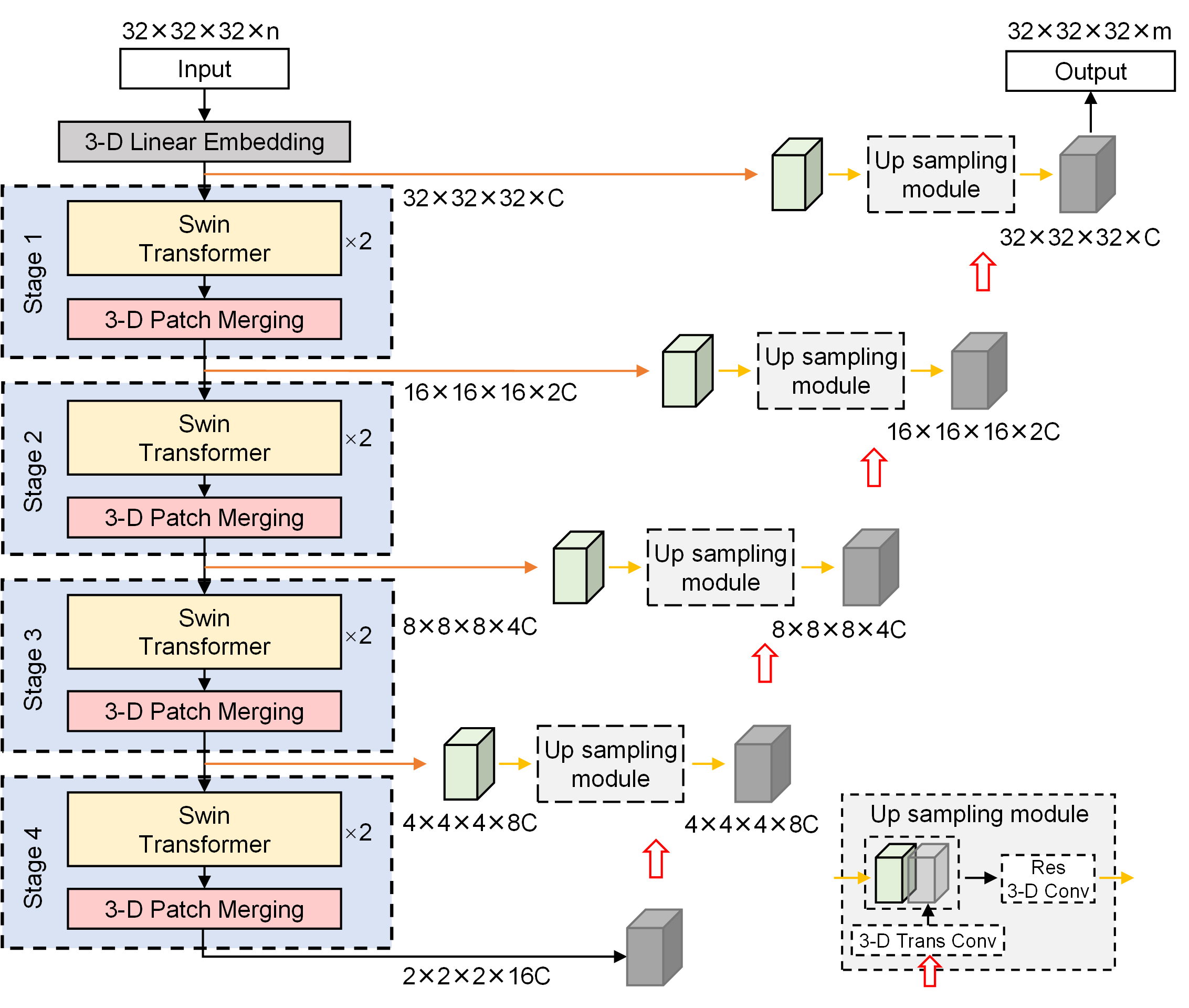}
	\caption{Network Architecture Diagram. $C$ denotes the number of channels of the feature map during network encoding and decoding, $n$ denotes the number of channels of the input data, and $m$ denotes the number of channels of the output data, depending on the forward modeling and inversion tasks.}
	\label{fig:fig2}
\end{figure}

\subsection{Network Architecture and Loss Functions}

The network used for the experiments in this paper is SwinUNETR\cite{hatamizadeh2021swin}~\cite{tang2022self}, which has been modified based on the characteristics of 3-D MT data\cite{wang2024three}. Due to its excellent performance in 3-D MT DL forward modeling, we consider it a representative network for application in the methodology proposed in this paper. As shown in Figure \ref{fig:fig2}, the network uses a 3-D Swin Transformer\cite{liu2021swin} as its core feature extraction module, incorporating a multi-level U-shaped encoding and decoding structure with skip connections to enhance contextual information interaction. During the encoding process, the number of feature map channels is initially increased through 3-D Linear Embedding to capture more abstract features. The network then sequentially passes through four down sampling stages, each containing two consecutive Swin Transformer modules and one 3-D Patch Merging module. The Swin Transformer module acts as the primary feature extraction component. Two consecutive Swin Transformer modules sequentially perform Window Multi-head Self-Attention (W-MSA) and Shifted Window Multi-head Self-Attention (SW-MSA) computations. MSA is based on Scaled Dot-Product Attention\cite{Vaswani2017Attention}, which can efficiently extract the feature weights from the data itself, and its calculation formula is as follows:

\begin{equation}
	{\rm{Attention}}({Q},{K},{V}) = {\rm{SoftMax}}(\frac{{{Q}{{K}^T}}}{{\sqrt d }}){V}
	\label{equ9}  
\end{equation}
where $Q$, $K$, and $V$ represent matrices composed of $query$, $key$, and $value$ vectors of sequential features, respectively; $d$ represents the dimension of the $key$ vectors. The window-based MSA is designed to enable self-attention computation on 2-D and higher-dimensional data, while the shifted window mechanism facilitates interaction between windows, thereby enhancing the network's receptive field. The 3-D Patch Merging module functions similarly to a pooling operation, reducing the size of the feature map while increasing the number of channels. Multi-stage down sampling allows the network to learn deep, abstract features from the data.

The decoding part of the network gradually restores the dimensions of the data to map them to the target output using the up sampling module. In the up sampling module, the deep feature maps are first enlarged using transposed convolution to match the size of the feature maps from the skip connections. These feature maps are then concatenated along the channel dimensions, followed by a convolution with a residual operation, before being passed to the subsequent up sampling module. Since the input data is normalized, the network applies a Sigmoid function to produce the final output. The number of input and output channels in the figure is determined by the specific task, with n = m = 1 for the forward modeling sub-network and n = 4, m = 1 for the inversion sub-network.

We use Mean Squared Error (MSE) as the loss function to guide network optimization during the training of each forward modeling sub-network and the overall network (training the inversion sub-network). For DL forward modeling, we trained separate forward modeling sub-networks for different forward responses with varying apparent parameters. The corresponding losses can be specified according to Equation (\ref{equ7}) as follows:

\begin{equation}
	{L_{fm}}(d,d'){\rm{ =  }}\frac{1}{n}{\sum\nolimits_{i = 1}^n {({d_i} - {{d'}_i})} ^2}
		\label{equ10}  
\end{equation}
where $n$ denotes the total number of data points, $d$ denotes the true value, and $d'$ denotes the network prediction result. Different apparent parameters (including ${\rho_{XY}}$, ${\phi_{XY}}$ , ${\rho_{YX}}$ and ${\phi_{YX}}$) can be denoted accordingly.

For the overall network training aimed at optimizing the inversion sub-network, forward modeling computations must be performed simultaneously using each forward modeling sub-network. Additionally, smoothing the network weights during training helps mitigate the overfitting problem\cite{krogh1991simple}, so we include a regularization term for the network weights, $L_w$. According to Equation (\ref{equ7}), (\ref{equ8}), and (\ref{equ10}), the overall network loss can be expressed as follows:

\begin{equation}
	\begin{array}{l}
		{L_{obt}} = \alpha ({L_{inv}} + {L_w}) + \beta \sum {{L_{fm}}} \\
		{L_{inv}} = \frac{1}{n}{\sum\nolimits_{i = 1}^n {({m_i} - {{m'}_i})} ^2}\\
		{L_w} = {\sum\nolimits_j W ^2}\\
		\alpha  = \frac{{{L_{inv}}}}{{{L_{inv}} + \sum {{L_{fm}}} }},{\rm{      }}\beta  = 1 - \alpha 
	\end{array}
\end{equation}
where $n$ denotes the total number of data points; $m$ represents the true value of the geoelectric model; $m'$ denotes the network's predicted value of the geoelecteric model; and $W$ represents the network weights. Adaptive adjustment strategies are applied to $\alpha$ and $\beta$. Since the forward modeling sub-networks are pre-trained, the value of $\sum {{L_{fm}}}$ will be smaller than ${L_{inv}}$, and ${L_{obt}}$ will be more heavily weighted in the optimization of the inversion sub-network.

\begin{figure}
	\centering
	\includegraphics[width=1\linewidth]{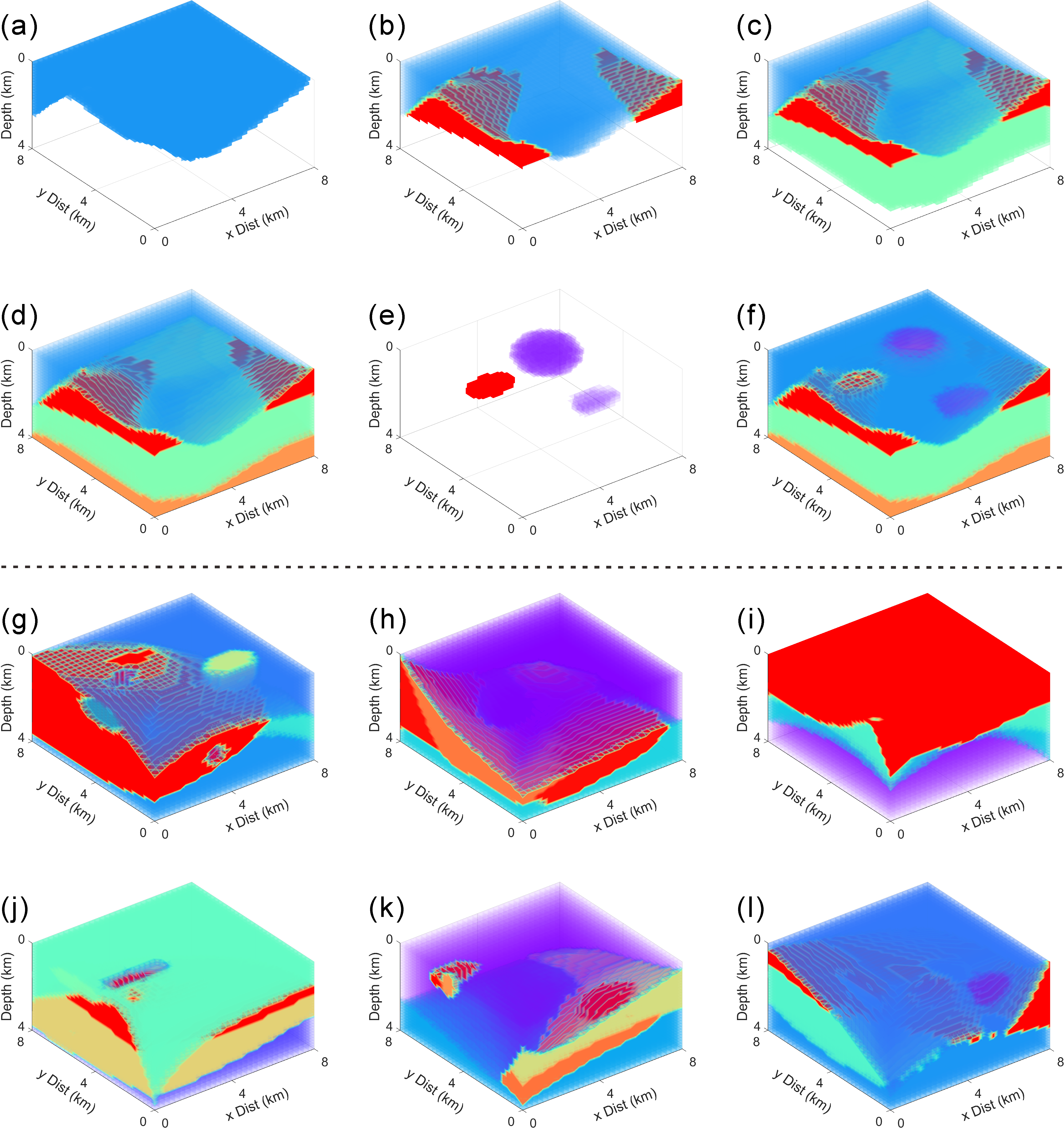}
	\caption{Schematic diagram of dataset production. (a) - (f) shows the generation process of the geoelectric model, and (g) - (l) shows part of the geoelectric model in the training set.}
	\label{fig:fig3}
\end{figure}

\section{Experimentation}
\subsection{Dataset Construction}

In order to explore the validity of the method and to make the network have the possibility of effective inversion of the field data, the resistivity distribution of the theoretical geoelectric model in the dataset of this paper simulates different morphologies of layers and regular anomalies at the same time, so as to make it close to the real subsurface tectonics. In the default Cartesian coordinate system, $x$ and $y$ denote the maximum distance in the horizontal direction of the geoelectric model, and $z$ denotes the maximum depth in the vertical direction, respectively. Each layer is generated by surface interpolation of a number of random 3-D spatial points, including four fixed points in the horizontal direction: (0, 0, $z'$), ($x$, 0, $z'$), (0, $y$, $z'$), and ($x$, $y$, $z'$), and several random coordinates $\left( {{p_1}, {p_2}, {\rm{z'}}} \right)$, where ${p_1} \in \left( {0, x} \right), {p_2} \in \left( {0, y} \right)$. The depth $z'$ of the spatial coordinates of each layer takes values between (0, $z$), with interleaving allowed for different layers. As shown in Figure \ref{fig:fig3}-a, the first layer $l_1$ is constructed and populated with random resistivity by the above method; then as shown in Figure \ref{fig:fig3}-b, we generate the second layer $l_2$, and when $l_2$ is deeper somewhere compared to $l_1$, the resistivity between the layers will be updated by the new random value, and vice versa will not be. Subsequently, from Figure \ref{fig:fig3}-c to Figure \ref{fig:fig3}-d, we gradually constructed a four-layer layered geoelectric model, which can simulate very diverse and complex subsurface formations because the layer surfaces formed by each random interpolation are allowed to overlap. In the second step, regular anomalies with random numbers, sizes and resistivities are added, mainly focusing on the shallow part, as shown in Figure \ref{fig:fig3}-e and Figure \ref{fig:fig3}-f, and finally a highly diversified theoretical geoelectric model including random layers and anomalies is obtained. Figure \ref{fig:fig3}-g through Figure \ref{fig:fig3}-i illustrate some of the geoelectric models in the training set.

Specifically, the resistivity of this dataset is randomly populated to take the range of (1, 1000) $ \Omega · m $. All geoelectric models are horizontally oriented at a distance of 8 km in x and y, uniformly divided into 32 × 32 equally spaced grids. The depth z is 4 km, with the first layer being 20m thick, divided into 32 grids using an isometric downward expansion factor of 1.1. The geoelectric model is forward modeling by FDM using ModEM\cite{Kelbert2014ModEM}, the simulated measurement points are uniformly distributed in the horizontal direction of the ground surface according to 32 × 32 with a spacing of 250m, and the simulated detection frequency is 1000Hz $\sim$ 0.1Hz, including 32 frequency points. Finally, the dimensions of the geoelectric model and its forward modeling response data with any apparent parameters are all 32 × 32 × 32.

Note that the model needs to be flared (5 grids in both horizontal and depth directions) due to forward modeling using ModEM. Therefore, we will actually generate a 3-D geoelectric model with a grid size of 42 × 42 × 37 for ModEM calculation of forward modeling data by the method described above, and dig down the first layer of its center to create a 3-D geoelectric model with a grid size of 32 × 32 × 32 for deep learning training, to avoid possible incorrect results of forward modeling calculations after simply filling the background field resistivity. We constructed 28,000 pairs of samples according to the above settings, which is a challenging dataset for both traditional inversion methods and DL inversion methods. Subsequently, they are randomly divided into training set, validation set, and test set according to 8:1:1, which are synchronously applied to the training of the forward modeling and inversion sub-networks.

\subsection{Data preprocessing}
To help the network learn the data features and to speed up the network fitting, we first normalize the data to reduce the differences in the range of values. Both the geoelectric modeling and forward modeling response data were mapped to between (0, 1). The normalization formula is as follows:
\begin{equation}
	{m_i}^\prime  = \frac{{{{\log }_{10}}({m_i}) - \min ({{\log }_{10}}(m))}}{{\max ({{\log }_{10}}(m)) - \min ({{\log }_{10}}(m))}},i = 1, \ldots ,N
\end{equation}
\begin{equation}
	{d{_i^\rho}}^\prime = \frac{{{{\log }_{10}}(d_i^\rho ) - \min ({{\log }_{10}}({d^\rho }))}}{{\max ({{\log }_{10}}({d^\rho })) - \min ({{\log }_{10}}({d^\rho }))}},i = 1, \ldots ,N
\end{equation}
\begin{equation}
	{d{_i^\phi}}^\prime = \frac{{d_i^\phi }}{{90}},{\rm{    }}i = 1, \ldots ,N
\end{equation}
where ${m_i}$, $d_i^\rho$, $d_i^\phi$ and denote each pre-normalized geoelectric model, apparent resistivity and phase, respectively. ${m_i}^\prime$, ${d{_i^\rho}}^\prime$ and  ${d{_i^\phi}}^\prime$ denote the post-normalization results, respectively. $N$ denotes the total number of datasets.

\begin{figure}
	\centering
	\includegraphics[width=1\linewidth]{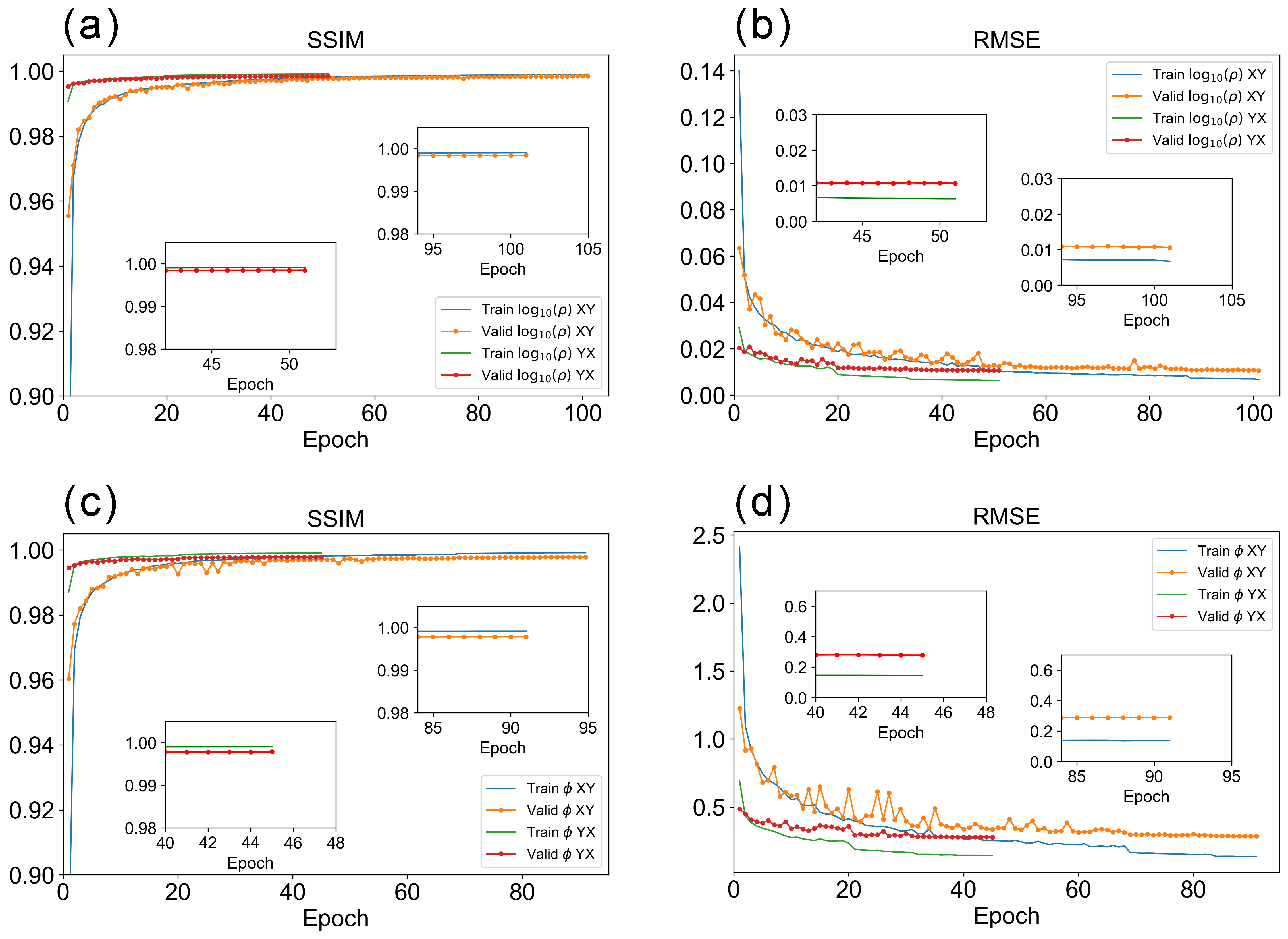}
	\caption{Variation curves of each metric in the training and validation sets of forward modeling sub-networks. (a) and (b) represent the SSIM and RMSE of the apparent resistivity, respectively; (c) and (d) represent the SSIM and RMSE of the phase, respectively.}
	\label{fig:fig4}
\end{figure}

\begin{table}[t]
	\caption{Forward modeling subnetworks test set results
		\label{tab:tab1}}
	\centering
	\linespread{1.25} \selectfont
	\begin{tabular}{ccccc}
		\hline
		& Log10(${\rho_{XY}}$) & ${\phi_{XY}}$    & Log10(${\rho_{YX}}$) & ${\phi_{YX}}$    \\ \hline
		SSIM & 0.9983     & 0.9977 & 0.9984     & 0.9977 \\
		MAE  & 0.0063     & 0.2003 & 0.0064     & 0.1945 \\
		RMSE & 0.0107     & 0.2959 & 0.0109     & 0.2878 \\ \hline
	\end{tabular}\\Note: The above metrics are all calculated in the data range (0, 1).
\end{table}

\subsection{Experimental environment and parameter settings}
The hardware environment for the experiment includes an INTEL i7-14700KF processor and an NVIDIA RTX 4090 D GPU, utilizing the Pytorch training framework and the RAdam optimizer. The batch size for forward modeling sub-network training is set to 24, with an initial learning rate of 2e-4 and a maximum of 100 training epochs. The batch size for overall training of the inversion sub-network is set to 8, with an initial learning rate of 2e-4 and a maximum of 65 training epochs. During training, if the validation set metrics do not improve for 5 consecutive epochs, the learning rate is reduced to half of its original value. If there is no improvement for 8 consecutive epochs, the early stopping mechanism is activated.
We quantitatively measure the accuracy of NN forward modeling and inversion by structural similarity (SSIM), mean absolute error (MAE) and mean mean square error (RMSE). Structural similarity (SSIM) focuses on the structural similarity of the data\cite{Hore2010SSIM}, which can reflect the local structural variations as well as the overall feature similarity of the data as follows:
\begin{equation}
	SSIM(x,y) = \underbrace {\frac{{(2{\mu _x}{\mu _y} + {C_1})(2{\sigma _{xy}} + {C_2})}}{{(\mu _x^2 + \mu _y^2 + {C_1})(\sigma _x^2 + \sigma _y^2 + {C_2})}}}_{{C_1} = {{({K_1}L)}^2},{\rm{  }}{C_2} = {{({K_2}L)}^2}}
		\label{equ15}  
\end{equation}
Here, $x$ denotes the true value; $y$ denotes the network predicted value; $\mu_x$ and $\mu_y$ denote the mean luminance; $\sigma_x$ and $\sigma_y$ represent the luminance variance. $\sigma_{xy}$ denotes the luminance covariance between the true value and the predicted value. $C_1$ and $C_2$ are constants used for stabilize the calculation and avoid division by 0. The variable $L$ in $C$ represents the range of variation of the data, for apparent resistivity and geoelectric modeling, $L$ is the range of values after taking logarithms, for phase, $L$ is 90. $K_1$ and $K_2$ are adjustment parameters used to control contrast and luminance, typically set to 0.01 and 0.03, respectively.

The MAE quantifies the overall deviation of the predicted values by calculating the mean of the absolute error between the predicted and actual values, as follows:
\begin{equation}
	MAE(x,y) = \frac{1}{n}\sum\limits_{i = 1}^n {(\left| {{y_i} - {x_i}} \right|)}
		\label{equ16}  
\end{equation}
where $n$ denotes the total number of data points in the data; $y_i$ denotes the predicted value of the network for the $i$-th data point; and $x_i$ denotes the actual value of the $i$-th data point.

The RMSE is more sensitive to larger errors and is used to identify the presence of extreme biases in the results, as follows:
\begin{equation}
	RMSE(x,y) = \sqrt {\frac{1}{n}\sum\limits_{i = 1}^n {{{({y_i} - {x_i})}^2}} }
		\label{equ17}  
\end{equation}
The parameters in Equation (\ref{equ15}) have the same meanings as those in Equation (\ref{equ16}). Additionally, we will assess the overall network inversion performance using 3-D data visualization.

\subsection{Forward modeling network training}

We initially train ${\rho_{XY}}$, ${\phi_{XY}}$ , ${\rho_{YX}}$ and ${\phi_{YX}}$ in the 3-D MT forward modeling response using four separate forward modeling subnetworks. In this process, ${\rho_{XY}}$ and ${\phi_{XY}}$ are fully trained first, and the resulting models are then used to train ${\rho_{YX}}$ and ${\phi_{YX}}$, respectively, using a transfer learning strategy to reduce the overall forward modeling training time. The variation curves of each metric in the training and validation sets are shown in Figure \ref{fig:fig4}. It can be observed that when dealing with the highly diverse layered 3-D geoelectric model presented in this paper, achieving high-precision forward modeling is more challenging compared to models containing only anomalous bodies, resulting in a certain degree of error in the simulation results. As the epochs increase, a slight overfitting phenomenon emerges: the training set metrics continue to improve, but the validation set metrics tend to stabilize. The results for each metric in the test set are shown in Table \ref{tab:tab1}, and they are consistent with the numerical results from the validation set. Subsequently, we use these four forward modeling sub-networks as fixed forward operators in the overall inversion network training and experimentally explore whether the DL inversion performance can be enhanced despite some forward modeling simulation errors.

\begin{figure}[t]
	\centering
	\includegraphics[width=1\linewidth]{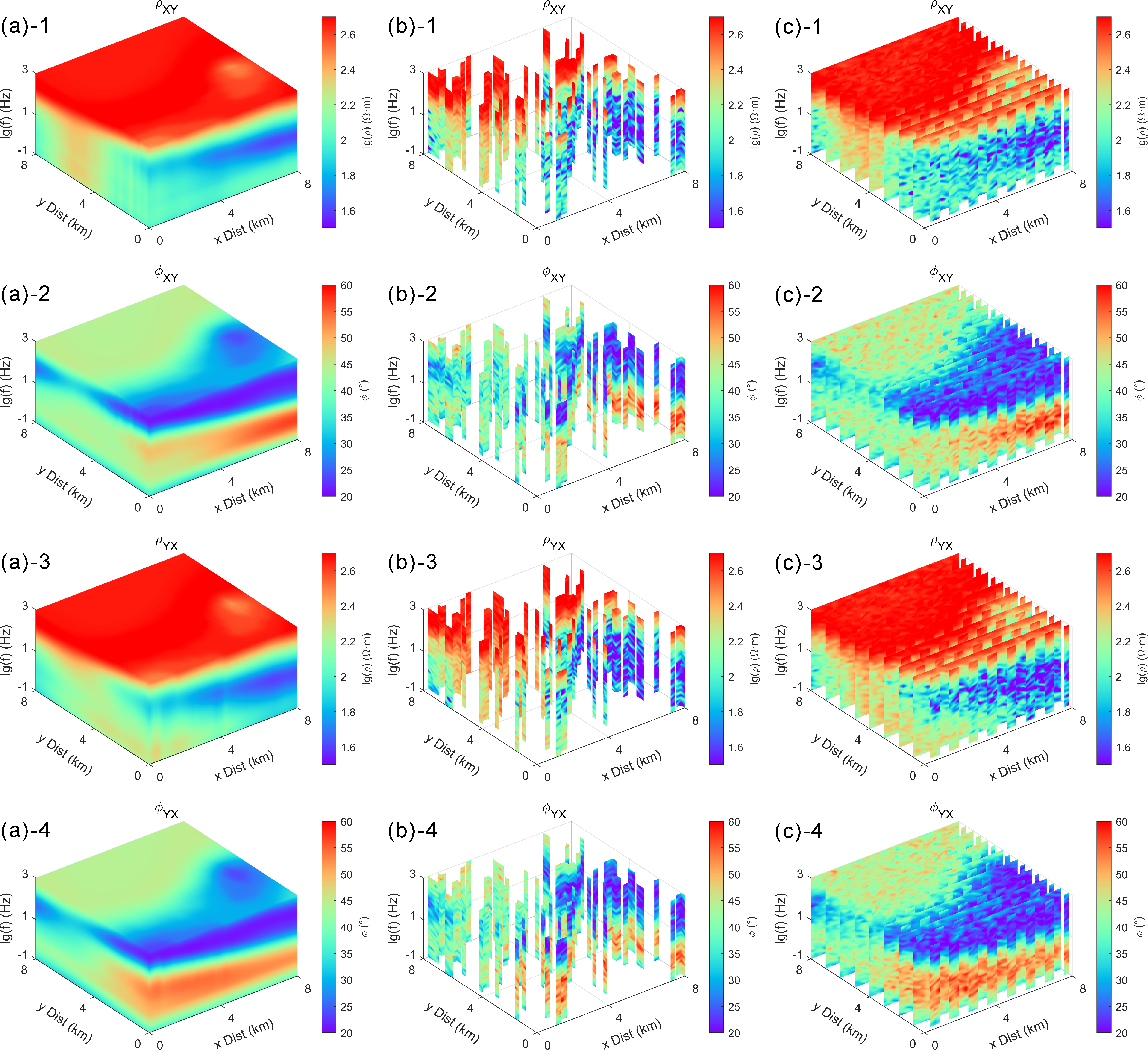}
	\caption{Input Data Mode. (a) Full data input mode, (b) Random mask \& noise-added input mode, (c) Regular mask \& noise-added input mode.}
	\label{fig:fig5}
\end{figure}
\begin{figure}
	\centering
	\includegraphics[width=0.7\linewidth]{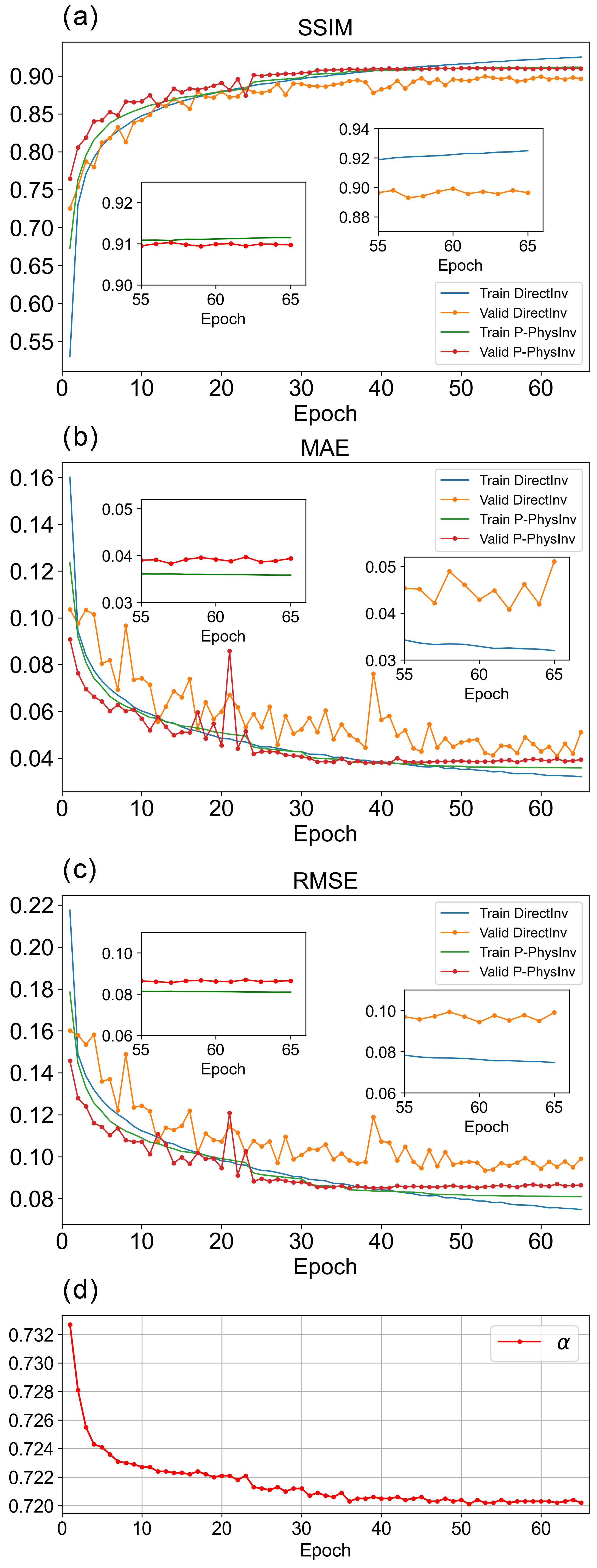}
	\caption{Variation curves of metrics for the training and validation sets of each inversion method. (a) SSIM variation curves (b) MAE variation curves (c) RMSE variation curves (d) $\alpha$ variation curves during training.}
	\label{fig:fig6}
\end{figure}
\begin{figure}
	\centering
	\includegraphics[width=1\linewidth]{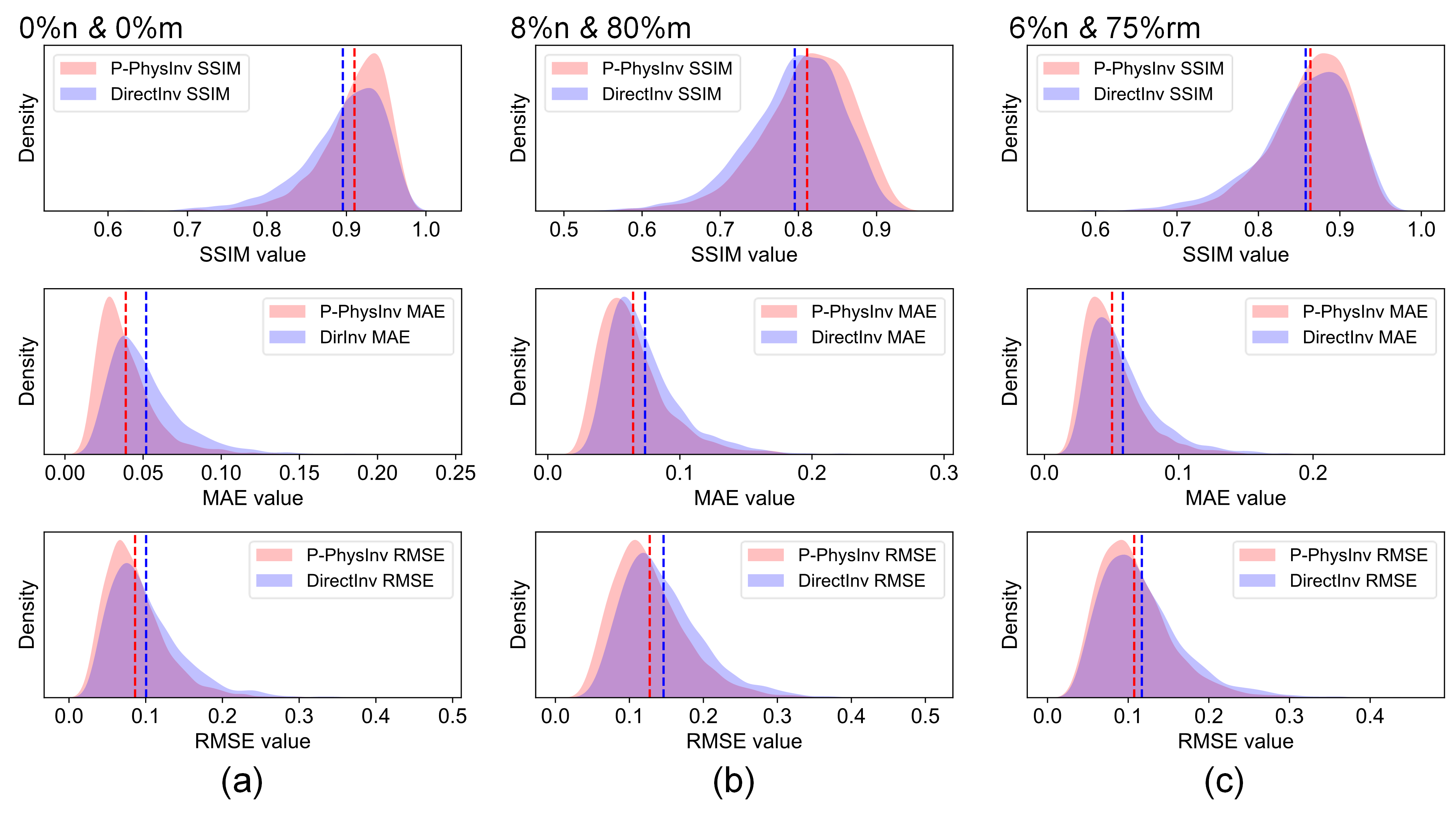}
	\caption{Probability density statistics for each metric in the test set of each inversion method. (a)	no noise and no masking input mode, (b) 8\% Gaussian noise with 80\% random masking input mode, (c) 6\% Gaussian noise with 75\% regular masking input mode.}
	\label{fig:fig7}
\end{figure}

\begin{table}[t]
	\caption{Results of test set metrics for each inversion method
		\label{tab:tab2}}
	\centering
	\linespread{1.25} \selectfont
	\begin{tabular}{cccc}
		\hline
		\multicolumn{4}{c}{Masking and Noise addtion Level}                                      \\ \hline
		\multicolumn{1}{c|}{} & 0\%n \& 0\%m     & 8\%n \& 80\%m   & 6\%n \& 75\%rm   \\ \hline
		\multicolumn{4}{c}{DirectInv log10($\rho$)}                                 \\ \hline
		SSIM                  & 0.8954           & 0.7953          & 0.8580           \\
		MAE                   & 0.0521           & 0.0736          & 0.0583           \\
		RMSE                  & 0.1005           & 0.1462          & 0.1169           \\ \hline
		\multicolumn{4}{c}{P-PhysInv log10($\rho$)}                        \\ \hline
		SSIM                  & 0.9105 (+1.7\%)  & 0.8142 (+2.4\%) & 0.8638 (+0.7\%)  \\
		MAE                   & 0.0384 (-26.3\%) & 0.0636 (-13.6\%)  & 0.0503 (-13.7\%) \\
		RMSE                  & 0.0852 (-15.2\%) & 0.1231 (-15.8\%)  & 0.1074 (-8.1\%)  \\ \hline
	\end{tabular}\\Note: The above metrics are all calculated in the data range (0, 1).
\end{table}

\subsection{Inversion network training}
Numerous studies have shown that DL inversion results significantly outperform traditional methods when applied to theoretical data. To evaluate the improvement of DL inversion with pseudo-physical information, we compare direct training inversion (DirectInv) with pseudo-physical information guided inversion (P-PhysInv). During training, we add 0-6\% Gaussian noise to the training set and mask 0-60\% of the measurement points to simulate the noisy, sparsely distributed data typically encountered in 3-D MT exploration (as shown in Figure \ref{fig:fig5}). Figure \ref{fig:fig6} illustrates the variation curves of metrics for the training and validation sets for each inversion method, and also shows the variation of the weight share $\alpha$ of the inversion sub-network in the loss function ${L_{obt}}$ during training. Table \ref{tab:tab2} presents the test set metrics for each inversion method, including results for cases with no noise and no masking, 8\% Gaussian noise with 80\% random masking, and 6\% Gaussian noise with 75\% regular masking. We tested more complex and sparse input data than those in the training set to evaluate the generalization ability of the methods and their potential applicability to field data.

As shown in Figure \ref{fig:fig6}, we trained the network for 65 epochs using both the DirectInv and P-PhysInv methods. As the epochs progressed, the validation set metrics for each method began to plateau, and different degrees of overfitting emerged. Notably, by the 50th epoch, DirectInv exhibited severe overfitting, with a significant gap between the training set metrics and the validation set metrics. In contrast, P-PhysInv significantly mitigated overfitting, resulting in a smaller gap between the training set and validation set metrics, with the validation set metrics overall outperforming those of DirectInv. Additionally, the weight share $\alpha$ of the inversion sub-network in the P-PhysInv method gradually decreased from 0.733 to 0.720, aligning with the trend observed in the validation set metrics. This suggests that the forward modeling sub-networks progressively increased their influence during training, helping to constrain the overall inversion direction and reduce the overfitting in DirectInv.

Table \ref{tab:tab2} summarizes the inversion metrics for the test set under three conditions: no noise and no masking, 8\% Gaussian noise with 80\% random masking, and 6\% Gaussian noise with 75\% regular masking. The results indicate that P-PhysInv outperforms DirectInv across all metrics. Specifically, in the case of no noise and no masking, P-PhysInv shows a 1.7\% improvement in SSIM, a 26.3\% reduction in MAE, and a 15.2\% reduction in RMSE compared to DirectInv. In the noisy and heavily masked case, P-PhysInv shows a 0.7\% to 2.4\% improvement in SSIM, an average reduction of approximately 14\% in MAE, and an 8\% to 16\% reduction in RMSE. Figure \ref{fig:fig7} illustrates the probability density of each metric in the test set across different levels of noise and masking. The figure shows that the P-PhysInv method primarily improves by reducing extreme biases and concentrating the high-precision inversion results. These experiments demonstrate that the P-PhysInv method, guided by pseudo-physical information, outperforms DirectInv. Additionally, the method remains effective in sparse and complex data environments, maintaining high evaluation metrics even when compared to the training set.

\begin{figure*}
	\centering
	\includegraphics[width=1\linewidth]{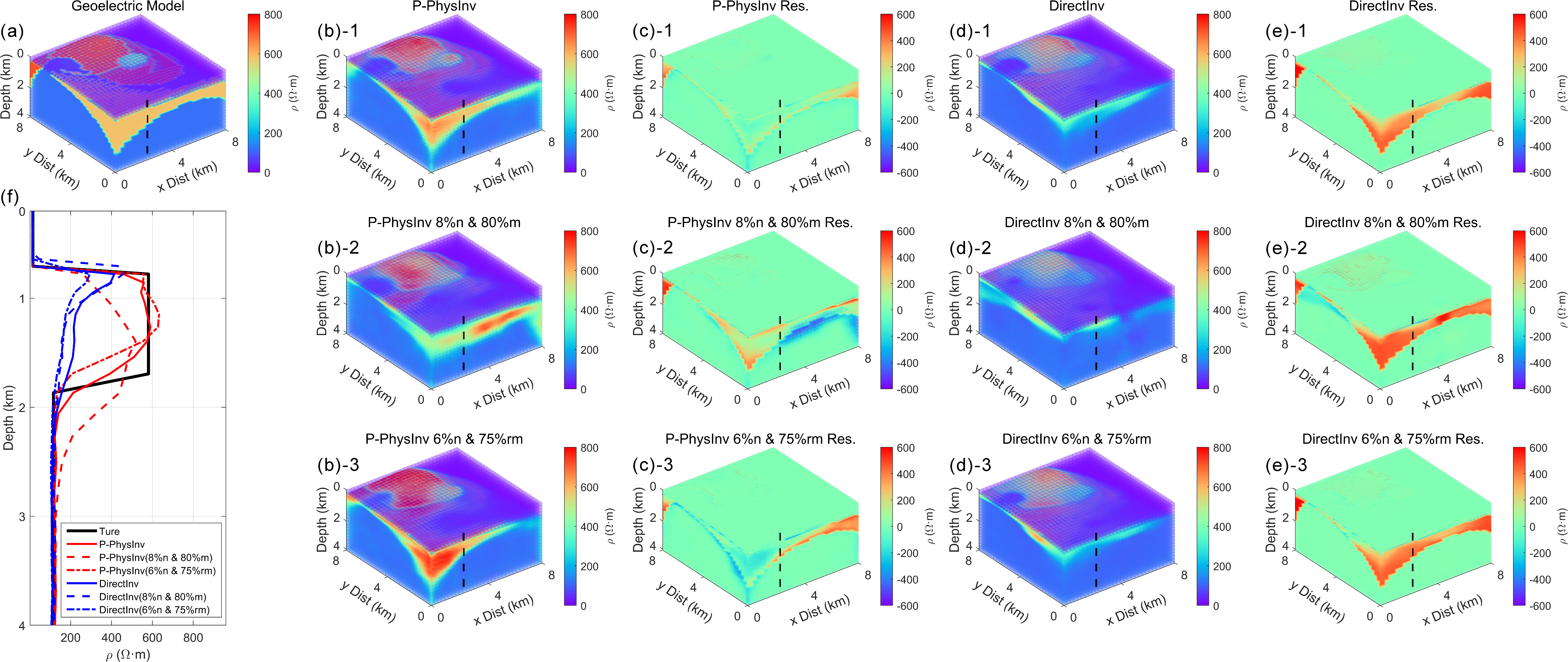}
	\caption{Comparison of inversion results for theoretical model - Case 1. (a) Actual geoelectric model, (b) P-PhysInv results, (c) Residuals of P-PhysInv results, (d) DirectInv results, (e) Residuals of DirectInv results, (f) Resistivity comparison curves for observations along the black dashed line. x\%n represents Gaussian noise at the x\% level, x\%m represents random masking at the x\% level, and x\%rm represents regular masking at the x\% level.}
	\label{fig:fig8}
\end{figure*}

\begin{figure*}
	\centering
	\includegraphics[width=1\linewidth]{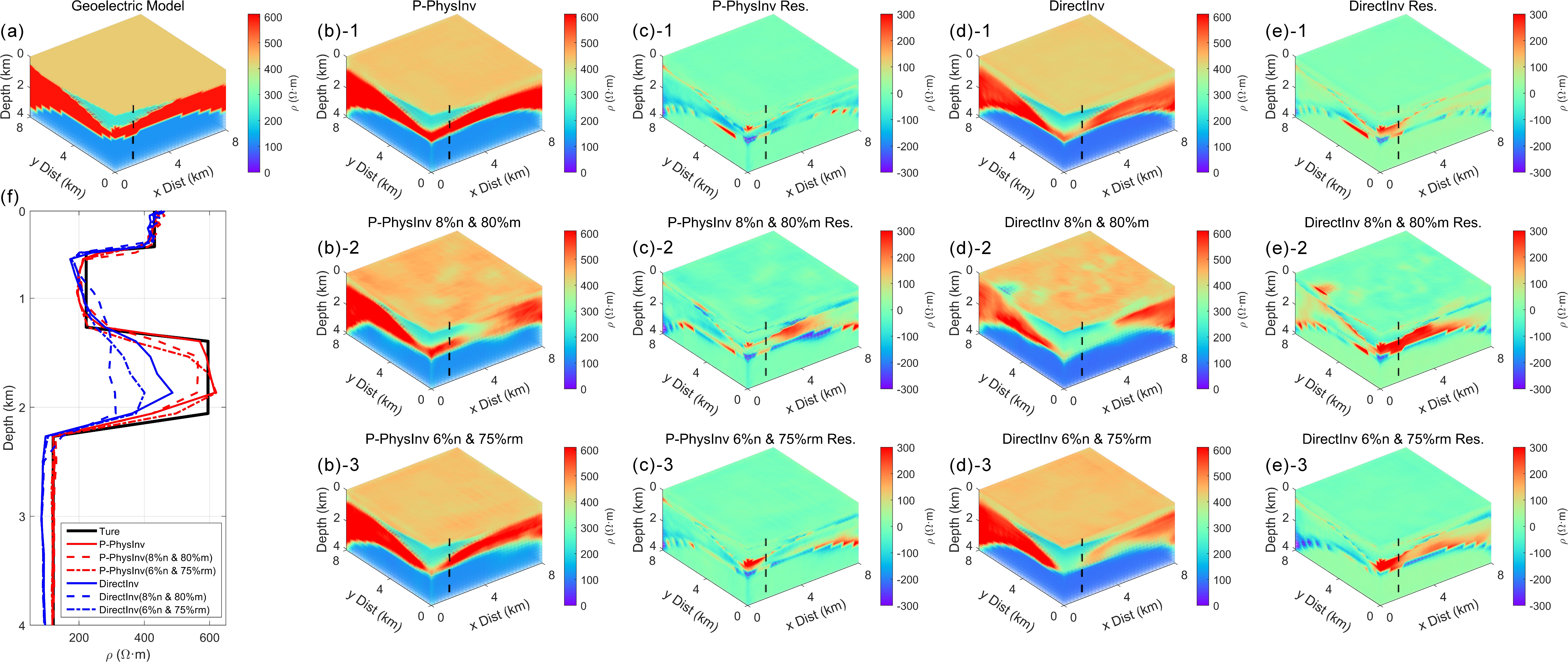}
	\caption{Comparison of inversion results for theoretical model - Case 2. (a) Actual geoelectric model, (b) P-PhysInv results, (c) Residuals of P-PhysInv results, (d) DirectInv results, (e) Residuals of DirectInv results, (f) Resistivity comparison curves for observations along the black dashed line. x\%n represents Gaussian noise at the x\% level, x\%m represents random masking at the x\% level, and x\%rm represents regular masking at the x\% level.}
	\label{fig:fig9}
\end{figure*}

\section{Discussion}
\subsection{Visualization comparison of different DL inversion methods}

We analyze the inversion results of different methods using 3-D visualizations. Figure \ref{fig:fig8} and Figure \ref{fig:fig9} display the inversion results of two randomly selected geoelectric models from the test set. In particular, Case 1, shown in Figure \ref{fig:fig8}, exhibits an overall low-high-low resistivity distribution, with a more complex distribution of multiple high resistivity layers in the center, necessitating higher inversion accuracy for the global structure. In terms of visualization, P-PhysInv provides more accurate inversion details and consistently outperforms DirectInv across various noise and masking levels. Particularly in the case of 8\% Gaussian noise with 80\% random masking, at the edge of the model (y = 0 km, x = 0 km to 8km), the DirectInv inversion performs poorly, making it nearly impossible to accurately reconstruct this portion of the high resistivity layer. Although the P-PhysInv method introduces some errors in the morphological details, it ensures that the overall structure is accurately represented. The resistivity variation curves shown in Figure \ref{fig:fig8}-f further demonstrate that the P-PhysInv method is more accurate in predicting the localization and resistivity values of different layers. Case 2, shown in Figure \ref{fig:fig9}, exhibits a high-low-high-low resistivity distribution near x = 0 km and y = 0 km, with a locally layered structure that requires precise portrayal of layer positions in the inversion results. The visualization results indicate that DirectInv performs poorly in resolving the interface between low and high resistivity from 1 km to 2 km; the layer interface appears blurred, and the depiction of layer orientation is inferior to that of the P-PhysInv results. The P-PhysInv method clearly outperforms DirectInv in multilayer geoelectric modeling with abrupt resistivity variations, particularly in accurately delineating the interfaces of high resistivity layers and the overall portrayal of irregular anomalies.

In summary, the DL inversion method effectively achieves high-precision and high-resolution inversion. For the more complex layered geoelectric model designed in this paper, the method demonstrates strong horizontal and vertical resolution, a capability that is difficult to achieve with traditional methods. In particular, compared to the DirectInv DL method, the P-PhysInv method, which incorporates pseudo-physical information, achieves better inversion accuracy across various noise and masking levels. This improvement is mainly observed in the more distinct inversion of layer interfaces and more accurate resistivity values. Additionally, we present test results at higher noise and masking levels compared to the training set, where the rule masks tested were not present during training, yet the method still proved effective. This demonstrates that the method is generalizable and has the potential for application to field data. It should be noted that although the overall implementation of the method is long (including about 5 days for dataset production, 35 hours for forward modeling training, and 28 hours for inversion training). However, the inference time of the trained inversion network for a single geoelectric model is only 0.016s, which is far more efficient than traditional inversion.

\begin{figure}[t]
	\centering
	\includegraphics[width=1\linewidth]{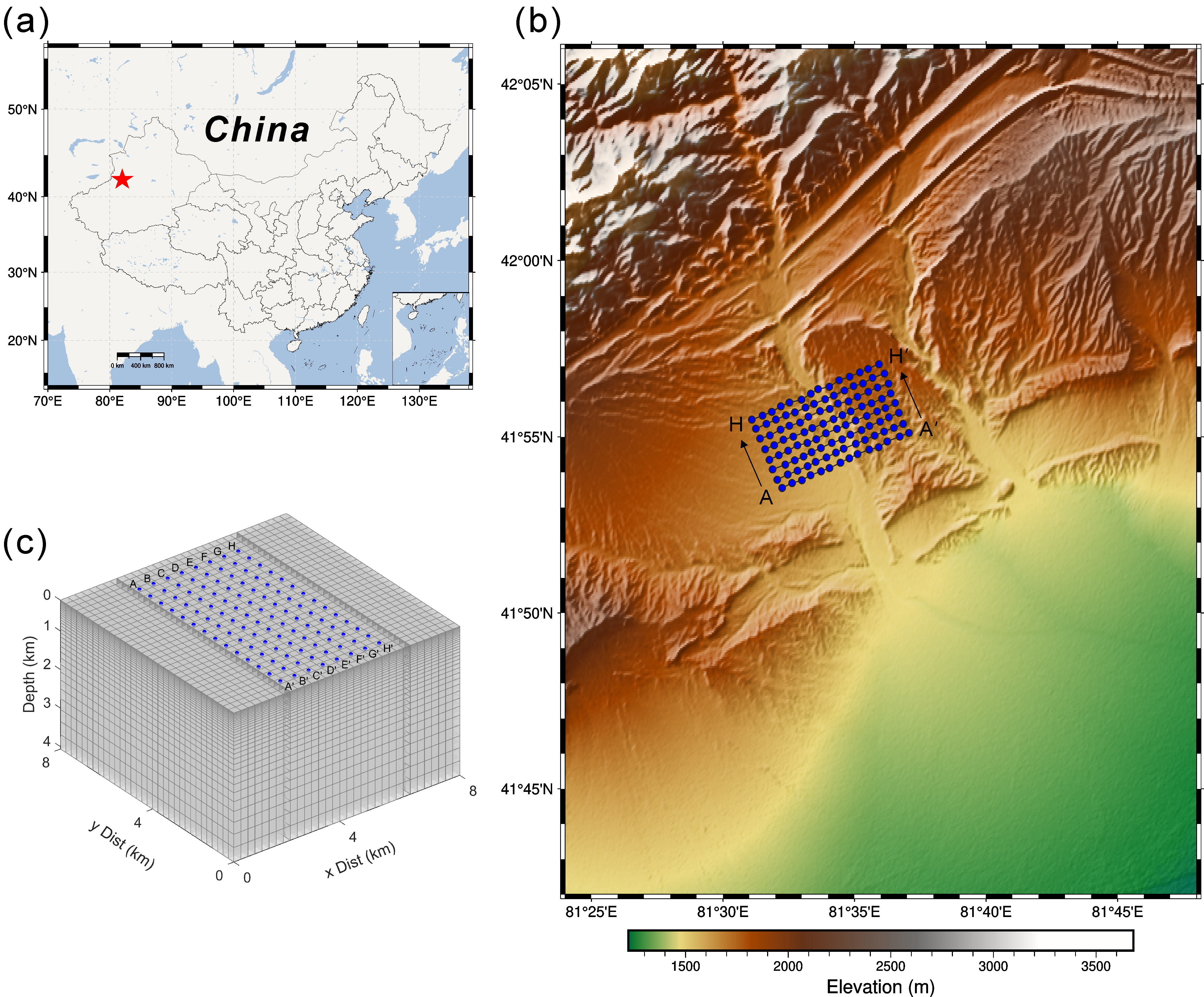}
	\caption{Overview of the work area. (a) Location of the work area; (b) Observation point deployment; (c) Schematic of DL inversion measurement point embedding.}
	\label{fig:fig10}
\end{figure}
\begin{figure}[t]
	\centering
	\includegraphics[width=1\linewidth]{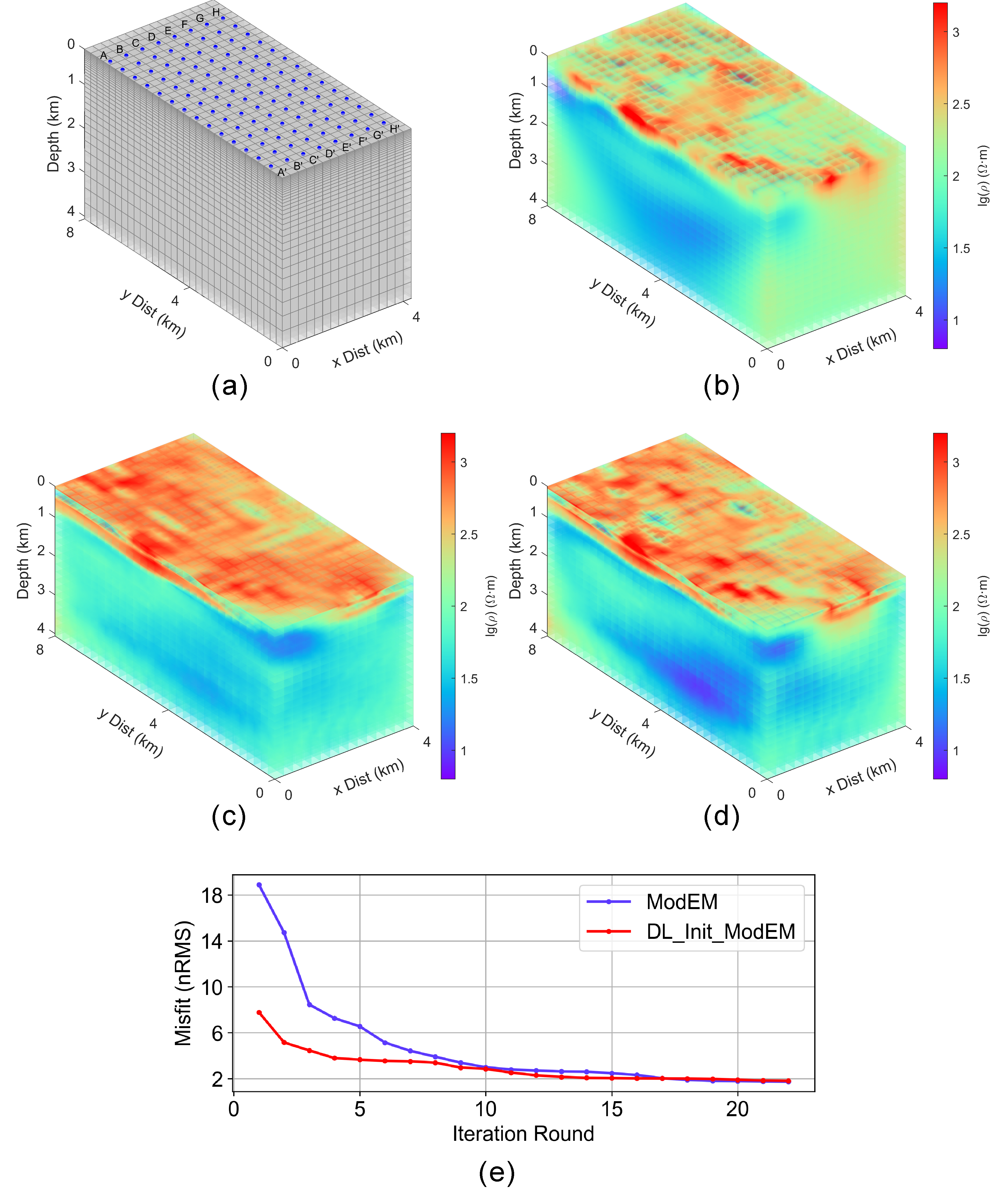}
	\caption{Field data inversion results. (a) Profile location; (b) ModEM inversion results; (c) DL end-to-end inversion results; (d) ModEM inversion results using DL end-to-end inversion as the initial model; (e) Variation curves of normalized root mean square misfit (nRMS)}
	\label{fig:fig11}
\end{figure}
\begin{figure*}
	\centering
	\includegraphics[width=1\linewidth]{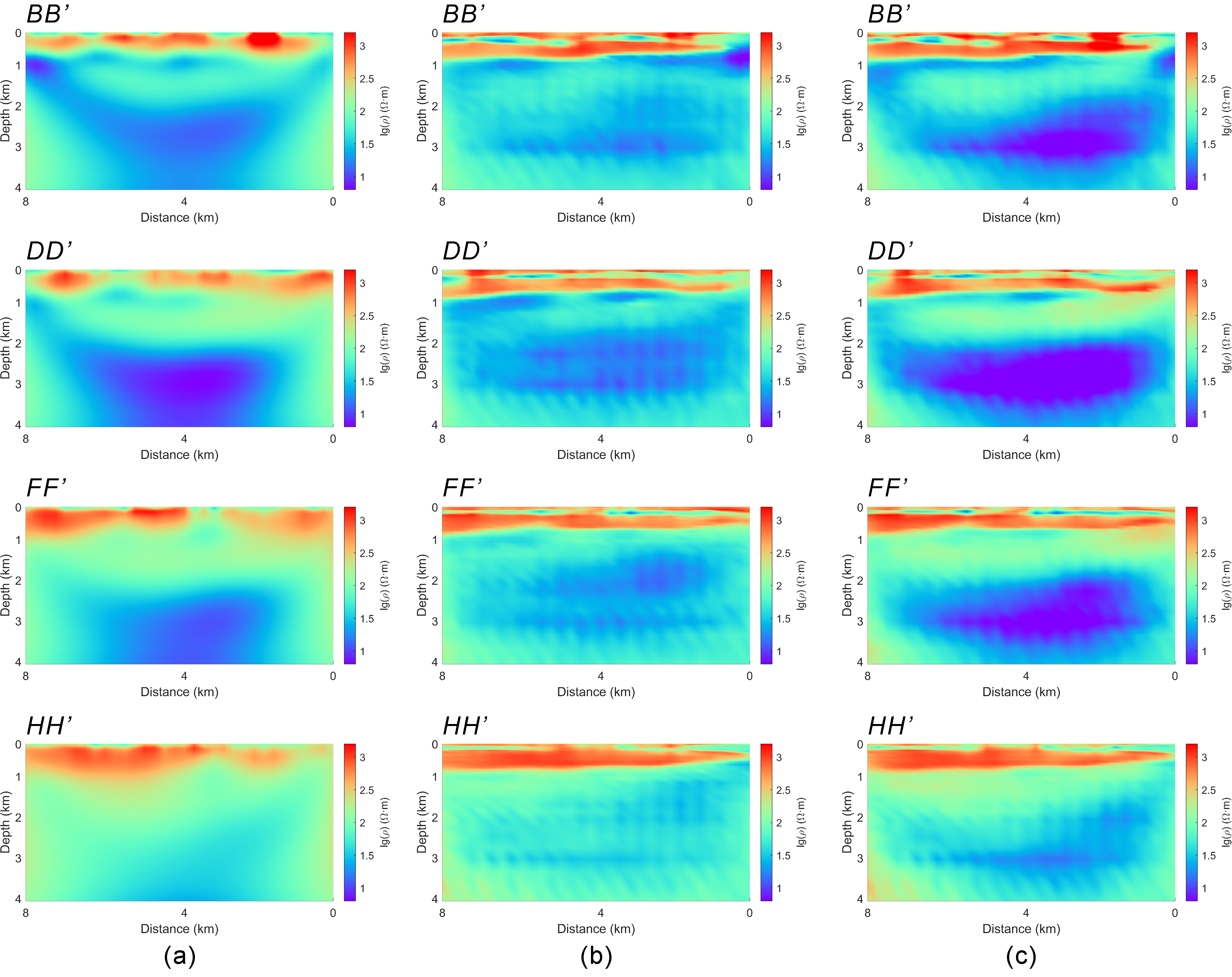}
	\caption{Field data inversion result profiles.(a) ModEM inversion results; (b) DL end-to-end inversion results; (c) ModEM inversion results using DL end-to-end inversion as the initial model.}
	\label{fig:fig12}
\end{figure*}

\subsection{Field data application}
By masking and adding noise to the data during training, this new input data mode simulates 3-D MT inversion under sparse data conditions, providing the flexibility to apply the method to 3-D MT field data. To evaluate the validity of the methods proposed in this paper on field data, we applied the P-PhysInv method to invert the 3-D MT data obtained from actual observations. As shown in Figure \ref{fig:fig10}, the work area is located in Xinjiang, China, where 112 measurement points are evenly distributed in 8 rows and 14 columns, with spacing between neighboring points ranging from 480m to 510m. The inversion is conducted using data from frequencies ranging between 320 Hz and 0.094 Hz. Due to the differences between the observed frequency points and the simulated frequency points in the theoretical dataset, the single-point observed data were first interpolated to 32 frequency points. The interval between each measurement point is larger than the simulated measurement point distance in the theoretical dataset. To maintain consistency in measurement point spacing, we uniformly embed the data into the 3D data structure, as shown in Figure \ref{fig:fig10}-c (blue dots). The non-data regions are set to -1 (considered as a masking operation, consistent with the theoretical model training), and the resulting 32 × 32 × 32 3-D data structure is input into the inversion network. For the network output, we extract only the data from the target region (the middle part) as the final result of the DL inversion.

Currently, there are two main approaches to applying DL inversion to field data: (1) training on a dataset that reflects the subsurface tectonic characteristics of the target work area or on a generalized dataset, followed by end-to-end inversion, and (2) performing conventional deterministic inversion using the end-to-end inversion results as the initial model. We tested both approaches, and the experimental results are displayed in Figure \ref{fig:fig11} and Figure \ref{fig:fig12}, with each profile in Figure \ref{fig:fig12} corresponding to the location indicated in Figure \ref{fig:fig11}-a. In Figure \ref{fig:fig11}-e, we show the variation of normalized root mean square misfit (nRMS) for ModEM inversion in the uniform half-space model and ModEM inversion using the DL end-to-end inversion results as the initial model. The ModEM inversion in the uniform half-space model reduces the misfit from 18.9 to 1.7 through more than 20 epochs. the DL end-to-end inversion has a misfit of 7.8, and after ModEM inversion using it as the initial model, the misfit gradually reduces to 1.8. It can be seen that the two have a similar final misfit, but the details of the inversion imaging are different. Visualization of the ModEM inversion in the uniform half-space model indicates that the surface in this region is characterized by high resistivity, with predominantly low resistivity up to z = 1 km, and localized low-high-low resistivity configurations from z = 1 to 3 km. The DL end-to-end inversion results are consistent with the ModEM inversion in the uniform half-space model results in terms of the general structure, showing high resistivity at the surface and low resistivity in the subsurface at a depth of 1 km, with clear vertical resolution. After using the DL end-to-end inversion results as the initial model in ModEM, the lateral resolution of high- and low- resistivity areas at the surface becomes more pronounced, and the low- resistivity areas at a depth of 2 km are also more prominent (as shown in the profiles BB’, DD’, and FF’ in \ref{fig:fig12}-c). The key difference from traditional inversion results is that the DL end-to-end inversion, along with the ModEM inversion results that use it as an initial model, also display strong layered and blocky structural characteristics. This is due to learning many theoretical models that have these characteristics. In addition, in order to make the dataset more generalizable, the theoretical geoelectric model is simulated over a much larger frequency range and therefore differs from the field data, which also affects the inversion results.

In summary, the inversion visualization of the field data indicates that the DL end-to-end inversion results are consistent with the ModEM inversion in the uniform half-space model results in terms of overall configuration, but the details differ significantly, with the DL method providing clearer layer resolution. After using the DL inversion results as an initial model for NLCG inversion, the details are further refined, resistivity is more accurately constrained, the misfit has also been further reduced, and the portrayal of interfaces in different resistivity regions becomes more distinct. Therefore, we believe that after training on a generalized dataset with subsurface tectonic guidance, the DL inversion method produces clearer interface features compared to traditional methods when applied to field data, demonstrating considerable practical potential.

\section{Conclusion}

In this study, we propose a pseudo-physical information-guided 3-D geo-electromagnetic (MT) deep learning (DL) inversion (P-PhysInv) method, which integrates pseudo-physical information from forward modeling into the training of the inversion network via neural networks (NNs) to guide its optimization. Theoretical experiments demonstrate that the P-PhysInv method outperforms the DL direct training inversion (DirectInv) method in handling complex theoretical geoelectric models with layered and irregular subsurface formations. The integration of pseudo-physical information enhances the accuracy of inversion results, particularly in delineating layer interfaces and resistivity, while significantly reducing network overfitting in complex datasets. Additionally, we simulate the real data environment of 3-D MT inversion during training and propose a new input data mode that includes masking and noise addition, enabling the method to be applied flexibly to real, sparse 3-D data. Field data inversion results indicate that the end-to-end inversion results produced by the P-PhysInv method under sparse data conditions are consistent with the ModEM inversion in the uniform half-space model results in terms of overall resistivity distribution, though they exhibit a strong training dataset signature. When the end-to-end inversion results are used as an initial model for deterministic inversion, the imaging details are further refined, demonstrating the method’s high practical potential.

\normalem
\bibliographystyle{IEEEtran}
\bibliography{ref}

\end{document}